\documentclass[aps, prb, reprint, amsmath, amssymb, groupedaddress, longbibliography]{revtex4-2}
\usepackage[colorlinks=true, allcolors=black]{hyperref}
\usepackage{graphicx}
\usepackage{xcolor}

\newcommand{\lt}{\left}
\newcommand{\rt}{\right}
\newcommand{\pa}{\partial}
\newcommand{\bk}{\mathbf{k}}

\newcommand{\bx}{\mathbf{x}}

\begin{document}

\title{Magnetic excitations of the strongly-anisotropic triangular XXZ model: a projected
spin-product state approach}

\author{Achille Mauri}

\author{Fr\'{e}d\'{e}ric Mila}

\affiliation{
Institute of Physics, Ecole Polytechnique F\'{e}d\'{e}rale de Lausanne (EPFL), CH-1015 Lausanne,
Switzerland}

\date{\today}

\begin{abstract}
Recent experiments have sparked interest on the ground-state properties and on the excitation
spectrum of the spin-$1/2$ triangular XXZ model with strong Ising-like anisotropy.
In this work, we introduce and analyze a ``projected'' coherent-state approximation to the model,
constructed by projecting spin-product states onto the low-energy configurations of the triangular
Ising antiferromagnet.
Within this approximation, we discuss the structure of the ground-state ordering and the
magnetization curve under a longitudinal field.
We then discuss the excitation spectrum by linearizing equations of motion deriving from a
time-dependent variational principle.
This leads to a natural description of spin-wave-like excitations within the strongly constrained
Hilbert space.
At zero field, the variational approximation leads to a very flat landscape of quasi-degenerate
states, which results in a low-energy pseudo-Goldstone mode in the excitation spectrum.
We compare our results to experimental measurements on the cobaltite K$_{2}$Co(SeO$_{3}$)$_{2}$
(KCSO).
Our approximation leads to a dramatic improvement as compared to linear spin-wave theory.
However, there remain deviations in the value of the pseudo-Goldstone gap and in the dispersion
near $M$ points at low field.
\end{abstract}

\maketitle

\section{Introduction}

The properties of spin systems with triangular lattice structures are a central theme in the field
of frustrated magnetism and, after decades of investigation, continue to be the subject of extensive
research.
Over the last years, a vast interest has been raised by new rare earth and transition metal
compounds with layered structures, in which the degrees of freedom governing the low-energy behavior
are effective spin $1/2$ arranged on triangular geometries~\cite{gallegos_prl_2025, li_sr_2015,
zhong_prm_2020, gao_npj_2022, arh_nm_2022, bag_prl_2024, scheie_prb_2024, gao_prb_2024,
zhu_prl_2024, zhu_npj_2025, xu_prb_2025, ulaga_prb_2025, ulaga_prb_2025b, woodland_prb_2025,
chen_nc_2026}.
Among the materials which were recently investigated, the cobaltite compound
K$_{2}$Co(SeO$_{3}$)$_{2}$~\cite{zhong_prm_2020, zhu_prl_2024, zhu_npj_2025, chen_nc_2026} (in the
following ``KCSO''), revealed a very strong easy-axis anisotropy, and an intriguing interplay
between a classical ``Ising-like'' frustration, and quantum mechanical effects.
More specifically, thermodynamic, magnetic, neutron diffraction, and inelastic neutron scattering
(INS) measurements~\cite{zhu_prl_2024, zhu_npj_2025, chen_nc_2026} showed that the exchange
interactions in this compound are described to a good approximation by a nearest-neighbour
spin-$1/2$ triangular XXZ model
\begin{equation} \label{H_exc}
H_{\rm exc} = \sum_{\langle i, j \rangle} \lt(J_{zz} S^{z}_{i} S^{z}_{j} + J_{xy}\lt(S^{x}_{i}
S^{x}_{j} + S^{y}_{i} S^{y}_{j}\rt) \rt)~,
\end{equation}
with a ratio $\alpha = J_{xy}/J_{zz}$ as small as $\alpha \simeq 0.07$.
The small value of $\alpha$ implies that the system, in a first approximation is controlled by the
Ising interaction $J_{zz}$, whereas $J_{xy}$ enters as a perturbation.

Despite the small values of $J_{xy}$, quantum effects have dramatic effects because of frustration.
Because the Ising exchange on the triangular lattice is geometrically frustrated, the $J_{zz}$ part
of the interaction has an infinite number of degenerate ground states.
These are the well known ``Wannier'' states~\cite{wannier_pr_1950}: the configurations in which
every triangular plaquette has either two ``up'' and one ``down'' spin, or two ``down'' and one
``up'' spin.
Due to this degeneracy, $J_{xy}$ has a controlling effect on the low-energy physics (at an energy
resolution of the order $J_{xy}$), determining how quantum effects lift the degenerate ground-state
subspace.

The low-energy Wannier states, as is well known, are in direct correspondence with hard-core dimer
coverings on the dual honeycomb lattice (there is a 2:1 correspondence: every dimer covering
corresponds to two Wannier states, related by an inversion of all spins).
The XXZ model in the small $J_{xy} \to 0$ limit, thus can be both viewed as a ``quantum Ising'' and
as a ``quantum dimer'' model~\cite{fazekas_pm_1974, moessner_prb_2001}.
The dimer model on the honeycomb lattice, in turn, realizes a $U(1)$ lattice gauge
theory~\cite{moessner_qdm_2011}.

The quantum-dimer, the quantum-Ising and, more generally, the XXZ model with finite ratio have been
a subject of investigations since decades~\cite{kleine_zpb_1992, kleine_zpb_1992b, burkov_prb_2005,
heidarian_prl_2005, wang_prl_2009, heidarian_prl_2010, ulaga_prb_2024, xu_prb_2025, zhu_npj_2025,
gallegos_prl_2025, ulaga_prb_2025, flores_calderon_prb_2025, hu_arxiv_2026}.
Theoretical approaches converge to the prediction that its ground state breaks translation symmetry,
and shows a spontaneous 3-sublattice order.
At zero and low magnetic field, different theoretical predictions point further to a
``spin-supersolid'' phase, in which $U(1)$ symmetry is broken alongside lattice-translation
invariance, although a different possibility was indicated recently~\cite{ulaga_prb_2025,
ulaga_prb_2025b}.

The recent experimental works on KCSO~\cite{zhu_prl_2024, zhu_npj_2025, chen_nc_2026,
zhu_arxiv_2026} have ignited a renewed interest on the properties of near-Ising triangular systems.
The observation of neutron scattering spectral functions has given direct access to low-energy
excitations (in the range $\epsilon \approx J_{xy}$), raising the challenge to understand
theoretically not only ground-state properties, but also the structure of excitations.

The experimental results reported in Refs.~\cite{zhu_prl_2024, zhu_npj_2025, chen_nc_2026,
zhu_arxiv_2026} were interpreted via a picture in which the ground-state at low fields is a
supersolid with ``Y'' structure, a state which has the same symmetry of the ground-configuration
predicted by a semiclassical (large-$S$) theory~\cite{miyashita_jpsj_1985, murthy_prb_1997,
kleine_zpb_1992}.
However, the spectral functions revealed features which cannot be explained simply in terms of
single-magnon excitations within linear-spin-wave theory (LSWT).
In particular, neutron experiments at low field revealed pronounced minima at the $M$ and $K/2$
points of the crystallographic Brillouin zone (referred to as ``roton-like'' minima in
Refs.~\cite{zhu_prl_2024, zhu_npj_2025, chen_nc_2026}), which are absent in the linear-spin wave
approximation.

Another feature revealed by neutron scattering measurements is the presence of excitations carrying
strong intensity at the $K$ point, and having small gap at small and moderate magnetic fields.
Linear-spin-wave theory for the XXZ model in a longitudinal field, with Hamiltonian $H_{\rm exc} -
\sum_{i} b S^{z}_{i}$, predicts a magnon mode at the $K$ point with gap $\epsilon_{\rm g}(K) \simeq
\sqrt{6 b J_{xy}}$ in the limit $J_{xy}/J_{zz} \to 0$.
The vanishing gap at $b = 0$ arises from an accidental degeneracy of the ground-state in the
classical approximation, which results in a soft mode in the dynamics of the system: a
``pseudo-Goldstone'' excitation.
It is natural to analyze the relation between the low-gap excitation observed experimentally at the
$K$ point, and a pseudo-Goldstone mode.
However, the gap of the pseudo-Goldstone mode at $b = 0$ vanishes only in the linear approximation,
while perturbative corrections beyond LSWT generate a gap~\cite{murthy_prb_1997, rau_prl_2018,
lin_arxiv_2025}.
In the case of a small ratio $J_{xy}/J_{zz}$ the $1/S$ expansion becomes highly nontrivial: in
the limit $J_{xy} \to 0$, the expansion parameter is effectively not $1/S$ but $J_{zz}/(S
J_{xy})$~\cite{kleine_zpb_1992, mauri_prb_2026}.
The first (one-loop) corrections to the gap, when calculated in the limit $1/S  \to \infty$,
$J_{zz}/(S J_{xy}) \to 0$ lead to a gap $\epsilon_{\rm g}(K, b=0) \simeq 3 \sqrt{S J_{xy}
J_{zz}}$~\cite{mauri_prb_2026}, which dramatically overestimates the experimental value.
In addition, the gap predicted by LSWT is strongly overestimated in comparison to neutron
measurements, for intermediate values of the field within the supersolid phase.

The spectrum of KCSO has been recently investigated by quantum Monte Carlo
simulations~\cite{zhu_npj_2025} based on the quantum dimer model related to the strongly
anisotropic XXZ interaction.
The results successfully reproduced the bottom of the neutron spectrum, confirming the validity of
the model as a description for the excitations in the compound, including the $M$-point minima.
Recent numerical investigations have analyzed ground-state properties and excitation spectra via
tensor-network approaches, both for KCSO~\cite{xu_prb_2025, hu_arxiv_2026}, and for other
anisotropic triangular systems~\cite{chi_prl_2022, gao_prb_2024, chi_prb_2024}, revealing complex
spectral features.

On the analytical side, different interpretations have been proposed, in terms of fractional spinon
excitations~\cite{jia_prr_2024, bose_prb_2025} or on the basis of mean-field
wavefunctions for the quantum dimer model and Schwinger-boson mean field
theory~\cite{flores_calderon_prb_2025}.

A natural alternative is to interpret the spectrum via spin-waves and to trace deviations from LSWT
to nonlinear effects and magnon-magnon interactions~\cite{chernyshev_prl_2006, chernyshev_prb_2009,
gallegos_prl_2025}.
However, constructing a systematic expansion beyond LSWT is difficult due to the strong coupling
nature of the problem at small $J_{xy}$~\cite{mauri_prb_2026}.

In the present work, we introduce an analogue to the classical approximation and to spin-wave
theory which is, by construction, well-defined in the highly-constrained quantum-Ising limit
$J_{zz} \to \infty$.
To this end, we construct and analyze variational wavefunctions obtained by projecting spin-product
states onto the subspace spanned by Wannier configurations.
The approach which we use is, in some sense, analogue to the method of Gutzwiller-projected
fermionic wavefunctions for the solution of the Hubbard or Heisenberg models~\cite{gros_ap_1989}.

By using the time-dependent variational principle (TDVP), restricted to the manifold of projected
spin-product states, we extend the approach to the study of the excitation spectrum, relevant to
understanding neutron scattering experiments.
In particular we consider a linear approximation of the TDVP equations at the variational
ground-state solutions.
This produces ``spin-waves'', consisting of small oscillations of local phases and magnetizations.

A variational study including correlations onto product-state wavefunctions was reported earlier in
Ref.~\cite{heidarian_prl_2010}.
Here, we consider simpler projected classical states, however, treating the angles before
projection as variational parameters, over which we optimize the energy.
In addition, here, we extend the analysis to include dynamical quantities, within a linear
time-dependent variational principle.

The article is organized as follows.
In Sec.~\ref{projected_spin_product_states} we introduce the projected spin-product states used as
variational wavefunctions throughout the rest of the work, and discuss the ``spin-wave'' equations
of motion which follow by linearizing the time-dependent variational principle about the ground
state.
In Sec.~\ref{application} we then apply this approach to the study of the XXZ model in a
longitudinal field.
First, we discuss the landscape of the variational energy at zero field, showing that it
presents a very flat direction (Sec.~\ref{zero_field_ground_state}).
In Sec.~\ref{sec_magnetization_curve} we then consider the ground state at finite field within the
supersolid phase $0 < b < b_{c} = 3J_{xy}/2$.
In Sec.~\ref{spectrum_finite_momentum}, the linearized time-dependent variational principle are
used to derive predictions for the excitation spectrum and the intensities.
Finally, in Sec.~\ref{excitation_spectrum_pG} we analyze in more detail the spectrum at the $K$
point, focusing on the pseudo-Goldstone mode.
We show that within the linearized time-dependent equations, a mode with low energy gap emerges
naturally from the flatness of the energy landscape.
The results of our analyses are compared to experimental data, reported for the KCSO compound, and
to semiclassical and linear spin-wave theory.
Sec.~\ref{conclusions} summarizes and concludes the article.

\section{Projected spin-product states}
\label{projected_spin_product_states}

In this work, we analyze a spin-$1/2$ system with Hamiltonian $H = H_{\rm exc} - b \sum_{i}
S^{z}_{i}$, where $H_{\rm exc}$ is the exchange interaction~\eqref{H_exc} and $b$ is a longitudinal
magnetic field.
In the Ising limit $J_{xy}/J_{zz}\to 0$, first-order perturbation theory allows one to reduce the
model to a projected theory~\cite{fazekas_pm_1974, wang_prl_2009}.
For $b < 3 J_{zz}/4$, the projected model is
\begin{equation}\label{H_proj}
H = J_{xy} \sum_{\langle i, j \rangle} P\lt(S^{x}_{i} S^{x}_{j} + S^{y}_{i}
S^{y}_{j}\rt)P - b \sum_{i} P S^{z}_{i} P~,
\end{equation}
where $P$ projects onto the subspace of triangular Ising ground states.
These are the well known ``Wannier states''~\cite{wannier_pr_1950}, satisfying the rule that all
triangular plaquette have 2-up-1-down or 1-down-2-up spins (``2U1D/2D1U rule'').

When $b > b_{c} =  3 J_{xy}/2$, the system maximizes the magnetization, and the ground-state is a
single ``up-up-down'' configuration (a $1/3$-magnetization plateau~\cite{zhu_prl_2024,
zhu_npj_2025, chen_nc_2026}).
Here, we focus on the region $0 < b < b_{c}$, where the system presents nontrivial quantum
fluctuations.

In this region, the 2U1D-2D1U rule makes it difficult to describe supersolid states in a
classical approximation for fixed $S = 1/2$.
In fact, in any classical state which satisfies 2U1D-2D1U rule exactly every canted spin must be
surrounded by six spins all of which are collinear to the $z$ axis.
A finite canting on adjacent sites, which is needed to gain transverse-exchange energy, is thus
suppressed in $J_{xy}/J_{zz}$.

To describe states in a spin-wave like fashion, we thus consider variational wavefunctions obtained
by projecting a classical state via the operator $P$ (a spin-product state projected to the
subspace of Wannier ground states).
The canting angles in this way can be arbitrary, while the 2U1D/2D1U rule remains satisfied exactly.

In the $S_{i}^{z}$ basis, the projected wavefunctions of the states which we consider are equal to
\begin{equation}\label{projected_wavefunction}
\psi_{\theta, \varphi} (\sigma_{1}, ...., \sigma_{N}) = Z_{\theta}^{-1/2} \prod_{i=1}^{N}
\Phi_{i}(\sigma_{i})~,
\end{equation}
for all spin configurations $\sigma_{1}, ..., \sigma_{N}$ satisfying the 2U1D/2D1U rule, and are
zero for all other configurations.
In Eq.~\eqref{projected_wavefunction}, $\Phi_{i}$ are single-site spin-coherent states, and
$Z_{\theta, \varphi}$ is a normalization factor determined by the condition $\sum_{\sigma}
|\psi_{\theta, \varphi}(\sigma_{1}, .., \sigma_{N})|^{2} = 1$.
The single-site coherent states $\Phi_{i}$ are conveniently parametrized in terms of the usual
spherical angles $\theta_{i}$, $\varphi_{i}$, as:
\begin{equation}
\Phi_{i} = \begin{pmatrix} 
            \Phi_{i}(\sigma_{i} = +1)  \\
            \Phi_{i}(\sigma_{i} = -1)
           \end{pmatrix}
= \begin{pmatrix}
   \cos\frac{\theta_{i}}{2} {\rm e}^{-i \tfrac{\varphi_{i}}{2}}\\
   \sin \frac{\theta_{i}}{2} {\rm e}^{i \tfrac{\varphi_{i}}{2}}
  \end{pmatrix}~.
\end{equation}

In this representation, $\theta_{i}$ and $\varphi_{i}$ represent the angles of the
spin $\mathbf{S}_{i}$ before projection (these angles can differ strongly from the angles
describing the direction of the physical average $\langle \mathbf{S}_{i}\rangle$, which is
calculated after projection).
The parametrization in terms of $\theta$ and $\varphi$ is natural because $Z$ and the averages of
products of of $S^{z}_{i}$ operators depend only on $\theta_{i}$ and not on $\varphi_{i}$.

To determine the state at zero, and finite field $b$ we use the variational principle, and
minimize the ground-state energy $\langle \psi |H|\psi \rangle = \langle \psi |PHP|\psi \rangle$ as
a function of $\theta_{i}$.
The probability distribution $|\psi (\sigma)|^{2}$ is equivalent to  that of a classical Ising
antiferromagnet with reduced Hamiltonian $\beta H = \sum_{\langle i, j\rangle} J_{zz}
S^{z}_{i}S^{z}_{j} - \sum_{i} h_{i} S^{z}_{i}$, in presence of a field  $h_{i} =\ln \tfrac{1 + \cos
\theta_{i}}{1 - \cos \theta_{i}}$, and in the limit $J_{zz} \to \infty$.
This allows us to calculate all relevant averages and correlation functions via a transfer-matrix
calculation, of the corresponding classical problem.
In particular, we used a transfer matrix calculation on cylinders with a finite width $L = 3n$ in
the $x$ direction, and with an infinite extent in the orthogonal $y$ direction.
The choice $L = 3n$ for the compact dimension was chosen in such way that the cylinders fit the
three-sublattice ordering.

In the projected XY model, and in the related quantum-dimer model, a crucial
result~\cite{wang_prl_2009} is that the Hamiltonians with $J_{xy} > 0$ and with $J_{xy}<0$
are unitarily equivalent: it is possible to explicitly construct a unitary transformation $U$ which,
acting on the projected Hilbert space, reverses the sign of $J_{xy}$.
In Ref.~\cite{wang_prl_2009} the transformation $U$ was constructed in the dimer language, but the
same construction can be directly converted to the spin language.
It is simple to see that the projected models with $J_{xy} > 0$ and $J_{xy} < 0$ models are
unitarily equivalent, both for $b=0$ and  for an arbitrary value of the longitudinal field $b$.
It follows that  models with opposite sign of $J_{xy}$ must have the same spectrum.

The manifold of projected coherent states however is not ``closed'' under the transformation $U$: if
$\psi$ is a projected coherent state, then $U \psi$ is not.
Thus, applying the projected coherent-state approximation to models with positive and negative
$J_{xy}$ leads to different variational energies.
It is simple to see that the variational approximation of the $J_{xy} < 0$ model has always
a lower energy.
The reason is that in the projected-coherent state approximation the transverse-exchange energy is
unfrustrated, and all angles $\varphi_{i}$ can be taken equal to $0$.
In the $J_{xy} > 0$ case, instead, the variational state has a frustrated transverse exchange, which
leads to a worse energy.
Due to unitary equivalence, this frustration is an artifact of the projected coherent-state
approximation: in reality the ground-state wavefunction has well defined signs determined by the
transformation $U$~\cite{fazekas_pm_1974, jiang_prb_2009, wang_prl_2009}.
For this reason, we apply the variational approximation only to the case $J_{xy} <0$, where it
does not introduce an artificial frustration, not really present in the exact solution to the model.

To analyze the antiferromagnetic case relevant to the cobaltite KCSO, we first perform the exact
unitary transformation $U$ to reverse the sign of $J_{xy}$ exactly, and then perform the projected
coherent state approximation to the resulting $J_{xy} < 0$ problem (similarly to
Ref.~\cite{wang_prl_2009}).

The variational approach just described leads to an energy of the form $E = - N J_{zz}/4 + N J_{xy}
f(b/J_{xy})$.
This energy provides a variational approximation not only to the projected model, but in
principle, also to the anisotropic XXZ model with an arbitrary ratio $J_{zz}/J_{xy}$.

\subsection{Linearized time-dependent variational principle and excitation spectrum}
\label{linearized_tdvp}

It is natural in the approach just introduced to consider ``spin-wave'' excitations consisting of
small oscillations of the local magnetizations $\langle S^{z}_{i}\rangle$ and the phases
$\varphi_{i}$.
In order to analyze these modes quantitatively, a natural description consists in applying the
time-dependent variational principle (TDVP)~\cite{cirac_rmp_2021} to the manifold of projected
spin-coherent states, and in linearizing the resulting equations about the energy minimum.

The TDVP equations for the problem can be expressed as the principle of stationarity of the  action:
\begin{equation} \label{S}
\begin{split}
S  & = \int {\rm d}t~ \langle \psi_{\theta, \varphi}| \lt( i \tfrac{\pa}{\pa t} - H \rt)
|\psi_{\theta, \varphi} \rangle\\
& = \int {\rm d}t~ \Bigg(\sum_{i} \langle S^{z}_{i} \rangle_{\theta(t), \varphi(t)}
\dot{\varphi}_{i} \\
& \qquad - E(\theta_{1}(t), \varphi_{1}(t), ..., \theta_{N}(t), \varphi_{N}(t))\Bigg)~.
\end{split}
\end{equation}

Here $|\psi_{\theta, \varphi}\rangle$ stands for a projected product state, with angles
$\theta_{1}$, $\varphi_{1}$, ..., $\theta_{N}$, $\varphi_{N}$, $\langle ... \rangle_{\theta,
\varphi}$ stand for averages within the state $|\psi_{\theta, \varphi}\rangle$, and $E = \langle H
\rangle_{\theta, \varphi}$.
To simplify the notation, we will use interchangeably $\langle ... \rangle = \langle ...
\rangle_{\theta, \varphi}$ below.

Linearizing the equations of motion which follow from the action $S$, and assuming a coplanar
ground state with all $\varphi_{i} = 0$ we find
\begin{equation} \label{eom1}
\begin{split}
2 &\sum_{j} C_{ij} \dot{x}_{j} + \sum_{j \in {\rm N.N.}(i)}  e_{ij} (\varphi_{i} -
\varphi_{j}) = 0~, \\
& \sum_{j} \lt( 2 C_{ij} \dot{\varphi_{j}} + L_{ij} x_{j} \rt) = 0~.
\end{split}
\end{equation}
where $x_{i} = \delta \theta_{i}/\sin \theta_{i}$, $C_{ij} = \langle (S^{z}_{i} - \langle S^{z}_{i}
\rangle ) (S^{z}_{j} - \langle S^{z}_{j} \rangle )\rangle$, $e_{ij} = J_{xy}\langle P(S^{x}_{i}
S^{x}_{j} + S^{y}_{i} S^{y}_{j})P\rangle$, and
\begin{equation}
\begin{split}
L_{ij} & = \sin \theta_{i} \frac{\pa}{\pa \theta_{i}} \lt(\sin \theta_{j} \frac{\pa}{\pa
\theta_{j}} E\rt) \\
& = \langle \{S^{z}_{i} - \langle S^{z}_{i} \rangle, \{H - E, S^{z}_{j} - \langle
S^{z}_{j} \rangle\}\}\rangle~.
\end{split}
\end{equation}

In this expressions, all averages are calculated in the variational ground-state. In particular,
$e_{ij}$ is the average exchange energy, within the variational state, of the bond connecting $i$,
$j$.

Using that $\sum_{j} C_{ij} x_{j} = -\frac{1}{2} \delta \langle S^{z}_{i}\rangle$, it can be seen
that Eqs.~\eqref{eom1} describe linear coupled oscillations of the phases $\varphi_{i}$ and of
the local magnetizations, thus corresponding to spin-wave-like modes.
Eq.~\eqref{S} has the same form as the action which governs the coherent-state representation of
spin models~\cite{auerbach_magnetism}, and its expansion near the energy minimum thus provides a
natural generalization of spin-wave theory.
The fact that the theory is projected onto the low-energy Ising manifold is encoded in the form of
the functions $\langle S^{z}_{i}\rangle$ and $E$, which are calculated within projected states.

The eigen-frequencies of the linear equations~\eqref{eom1} are given by the eigenvalues of the
dynamical matrix
\begin{equation} \label{eom2}
D =  \frac{i}{2} \begin{vmatrix}
      0_{N \times N} & -C^{-1} L C^{-1}  \\
      M & 0_{N \times N} \\
     \end{vmatrix}~,
\end{equation}
where $M_{ij} = \delta_{\langle i, j \rangle} e_{ij}- \delta_{ij} \sum_{k} e_{ik}$ and
$\delta_{\langle i, j \rangle}$ is the nearest-neighbour adjacency matrix.

In states with 3-sublattice order, which we consider here, the spectrum of $D$ gives rise to three
dispersion branches (``magnon'' modes), one for each atom in the magnetic unit cell.

To compare the results of the linearized TDVP approximation to experimental spectra, it is useful to
compute not only the frequencies, but also the contribution of the three branches to a dynamical
structure factor.
Here, we analyze in particular the $S^{z}$ correlation function ${\cal S}^{zz}(\bk, \omega)
= N^{-1} \int_{-\infty}^{\infty} {\rm d}t \sum_{i, j} \langle 0|S^{z}_{i}(t) S^{z}_{j}(0)|0 \rangle
{\rm e}^{-i \bk \cdot(\bx_{i} - \bx_{j}) + i \omega(t-t')}$.
This is particularly relevant as longitudinal terms were indicated as dominant contributions to
inelastic neutron scattering in KCSO~\cite{zhu_npj_2025, zhu_arxiv_2026}, at least at low fields.
To determine ${\cal S}^{zz}(\bk, \omega)$ in a framework consistent with the approximations above,
we consider the linear response to a source term $H_{J}(t) = - \sum_{i} J_{i}(t) S^{z}_{i}$ added to
the Hamiltonian.
The response function $\chi^{zz}_{ij}(t-t') = \delta \langle S^{z}_{i}(t)\rangle/\delta J_{j}(t')
\big|_{J = 0}$ gives access to ${\cal S}^{zz}$ via the fluctuation-dissipation relation ${\cal
S}^{zz}(\bk, \omega) = \frac{1}{\pi} \Theta(\omega) {\rm Im} \chi^{zz}(\bk, \omega)$.
Within the linearized TDVP approximation, we compute $\chi$ by calculating the infinitesimal
perturbation to the trajectories $\theta_{i}(t)$, $\varphi_{i}(t)$ induced at linear order in
$J_{i}(t)$.
By an explicit calculation, illustrated in Appendix~\ref{dynamical_structure_factor}, we derive the
matrix elements $|\langle n, \bk |S^{z}_{-\bk}|0\rangle|^{2} =  \frac{1}{N} \lt| \sum_{i} {\rm e}^{i
\bk \cdot \bx_{i}} \langle n, \bk |S^{z}_{i} |0\rangle \rt|^{2}$ in terms of simple expressions,
involving eigenvectors and eigenvalues of $D_{\bk}$.

\section{Application to the strongly-anisotropic triangular XXZ model}
\label{application}

\subsection{Ground state at zero field. Quasi-degenerate states}
\label{zero_field_ground_state}

As discussed above, we apply our variational approximation to projected models with $J_{xy} < 0$,
for which there is no frustration of the transverse exchange interaction.
We thus consider coplanar states in which all $\varphi_{i} = 0$.
Assuming three sublattice ordering, we consider as variational parameters the three bare angles
$\theta_{\rm A}$, $\theta_{\rm B}$, $\theta_{\rm C}$ for spins on the three sublattices A, B, C.

For zero field, we calculated the variational energy $E$ via the transfer matrix method on a
cylinder of width $L = 18$.
After minimization, we find that the energy minimum occurs at $\theta_{\rm A} = \theta_{\rm B} =
0.48785(5) \pi$, $\theta_{\rm C} = 0.5454(3) \pi$, and has energy $E_{xy} = E-E_{\rm Ising} = -
0.139695(5) N J_{xy}$.
(The uncertainties are estimated by comparing the results to those obtained  on cylinders of width
$L=15$).
Calculating the average magnetizations in this variational state we find $\langle \mathbf{S}_{\rm
A}\rangle = \langle \mathbf{S}_{\rm B} \rangle = (0.19714(4), 0,  0.2218(1))$, $\langle
\mathbf{S}_{\rm C} \rangle = (0.0725(1), 0,  -0.4298(1))$ for the projected model with $J_{xy}<0$
(see Appendix~\ref{averages}).

For the model with antiferromagnetic $J_{xy}$, we take the variational ground state to be $U \psi$,
where $U$ is the unitary transformation which reverses the sign of the transverse exchange.
The magnetizations, therefore, have to be calculated by undoing the unitary mapping
$U$~\cite{wang_prl_2009}.
We find that the ground-state average magnetizations are $\langle \mathbf{S}_{\rm A} \rangle =
(0.19450(3), 0,  0.2218(1))$, $\langle \mathbf{S}_{\rm B}\rangle = (-0.19450(3), 0, 0.2218(1))$,
$\langle \mathbf{S}_{\rm C}\rangle = (0, 0, -0.4298(1))$.
The ground-state predicted variationally is therefore a ferrimagnet with a small net moment in the
$z$ direction, consistently with predictions of the earlier variational analysis in
Ref.~\cite{sen_prl_2008, wang_prl_2009}.
In the AFM case, this ferrimagnetic state has the ``Y'' structure (the three magnetization vectors
on the three sublattices are arranged as the three segments composing the letter ``Y'').
The same structure was found in the earlier analyses of the easy-axis XXZ model reported in
Refs.~\cite{melko_prl_2005, wang_prl_2009, heidarian_prl_2010, yamamoto_prl_2014, xu_prb_2025}.

Although the ferrimagnetic state gives the minimal energy, we find that the energy landscape has a
very soft direction, along which the variational energy is almost flat.
This line interpolates continuously between the ferrimagnetic state and a competing
antiferromagnetic (UD0) solution~\cite{miyashita_jpsj_1985, kleine_zpb_1992, murthy_prb_1997,
burkov_prb_2005}, which has zero net moment in the $S^{z}$ direction.
After optimization, we find that the antiferromagnetic solution has variational angles $\theta_{\rm
A} = \pi - \theta_{\rm B}=0.4704(2) \pi$, $\theta_{\rm C} = \pi/2$ and energy $E_{xy} = -0.13958(1)
N J_{xy}$, which differs from the energy of the ferrimagnetic state by less than $1/1000$.
For the average moments in the UD0 state we find via the transfer matrix method $\langle
\mathbf{S}_{\rm A}\rangle = (0.1127(2), 0, 0.3732(5))$, $\langle \mathbf{S}_{\rm B} \rangle =
(0.1127(2), 0, -0.3732(5))$, $\langle  \mathbf{S}_{\rm C}\rangle = (0.2400(5), 0, 0)$ in the
$J_{xy} < 0$ case and $\langle \mathbf{S}_{\rm A}\rangle = (-0.094688(4), 0, 0.3732(5))$, $\langle
\mathbf{S}_{\rm B} \rangle = (-0.094688(4), 0, -0.3732(5))$, $\langle \mathbf{S}_{\rm C}\rangle =
(0.2392(4), 0, 0)$ in the $J_{xy}>0$ case.

\begin{figure}
\centering 
\includegraphics[scale=1]{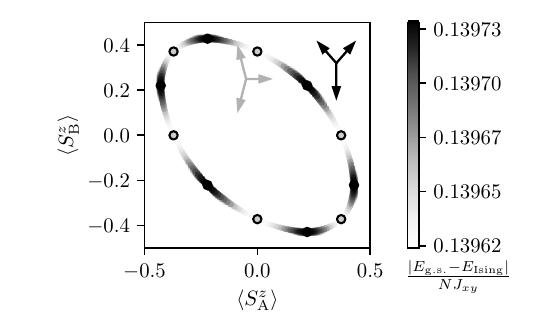}
\caption{\label{landscape} Line of quasi-degenerate projected states at zero longitudinal field for
$L=12$.
The arrows show the sublattice magnetizations in the AFM case in the Y state (black) and in the UD0
state (grey), which are respectively local minima and maxima of the variational energy along the
line.}
\end{figure}

\begin{figure}
\centering
\includegraphics[scale=1]{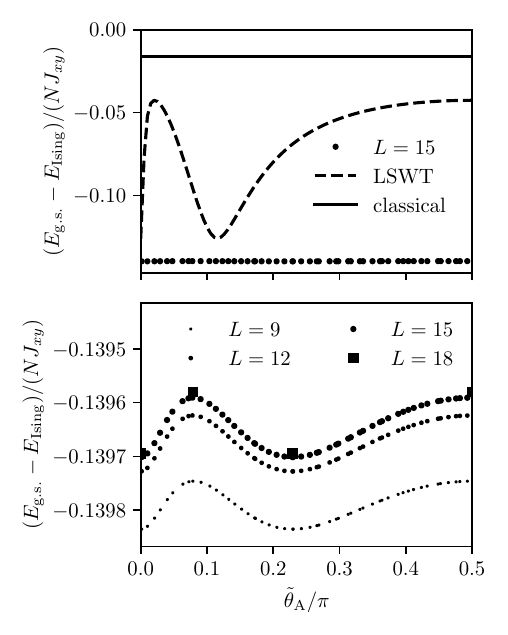}
\caption{\label{landscape2} Energy as a function of the canting angle at sublattice A.
Solid and dashed lines show, respectively, the classical energy and the classical + $1/S$ energy
along the classically-degenerate ground-state manifold for $J_{xy}/J_{zz} = 0.07$.
The dots show the minimum energy of a projected spin-product state with fixed canting angle
$\tilde{\theta}_{\rm A} = \arctan (\langle S^{x}_{i} \rangle/\langle S^{z}_{i} \rangle)$.
The bottom panel enlarges the narrow interval of energies separating the Y and the UD0 solutions,
and shows results obtained by the projected spin-product approximation for different $L$.
}
\end{figure}

To determine in more detail the structure of the energy landscape we generated solutions near the
minimum by minimizing with respect to $\theta_{\rm C}$ for fixed values of $\theta_{\rm A}$ and
$\theta_{\rm B}$.
We then analyzed the minimal energy for fixed values of the canting angle of the A-sublattice atoms
$\tilde{\theta}_{\rm A} = \arctan (m^{x}_{\rm A}/m^{z}_{\rm A})$.
We find that, in all low-energy solutions, all bare angles $\theta_{\rm A}$, $\theta_{\rm B}$,
$\theta_{\rm C}$ are quite close to $\pi/2$.
However, the physical magnetizations change widely, producing a soft line of quasi-degenerate
states with physically distinct ground-state averages (see Fig.~\ref{landscape},~\ref{landscape2}).

The Y and UD0 states are respectively local minima and maxima along the quasi-degenerate line, and
for all points on the manifold, the variational energy changes by less than $1/1000$.

The approximate degeneracy which we find has the same structure as the well-known degeneracy of
the classical XXZ model at zero field.
The latter has a one-dimensional manifold of exactly degenerate ground
states~\cite{miyashita_jpsj_1985, kleine_zpb_1992, murthy_prb_1997, gao_prb_2024}, which also
interpolates between ``UD0'' and ``Y'' states.
The states which we find in our variational wavefunction approach, however, are very different
from the classical ones.
In the classical ground states, when we take the limit $J_{xy}\to 0$ all spins become almost
collinear to the $z$ axis, and the energy behaves as $E = E_{\rm Ising} - N S^{2}
O(J_{xy}^{2}/J_{zz})$~\cite{fazekas_pm_1974}.
In the limit in which $J_{xy} \to 0$, however, the exact energy of the model must \emph{a priori}
behave as $E -E_{\rm Ising} =  O(J_{xy})$ since in this limit $J_{xy}$ can be factorized as a
global energy scale.
The transverse exchange energy of the classical approximation is thus 0 in the relevant energy
range (linear in $J_{xy}$).
The states considered here, instead, have finite canting angles and a correct energy-dependence $E
= E_{\rm Ising} - N O(J_{xy})$.
The numerical value of the energy which we find $\simeq - 0.1397 N J_{xy}$ is a significant
fraction of the energy $E_{xy} = -(0.16 \pm 0.005) N J_{xy}$ which was deduced in
Ref.~\cite{kleine_zpb_1992b} based on extrapolations of exact diagonalization results.

This shows that, surprisingly, the classical prediction of degenerate states remains almost valid
after projection, despite the projector leads to a drastic modification of the states.

In the semiclassical approximation, the degeneracy becomes lifted by the linear-spin-wave vacuum
energy. It is known that this effect favors the ``Y'' configuration, which has the lowest spin-wave
energy among the classical vacua~\cite{kleine_zpb_1992, murthy_prb_1997, rau_prl_2018}.
In the linear-spin-wave approximation, the lifting of the degeneracy induced by the spin wave
energy is strong for $J_{xy} \to 0$.
This is illustrated in Fig.~\ref{landscape2} (dashed line) in the case $J_{xy}/J_{zz} = 0.07$:
as it can be seen the spin-wave energy produces a very deep well.

In the projected approach which we considered, the variational wavefunctions already include a
significant amount of quantum fluctuations but, still, can be considered as analogue to
``classical'' states, living within the constrained Hilbert space.
In other words, they do not include the energy of spin-wave vacuum fluctuations, which in the $1/S$
expansion are responsible for the lifting of the degeneracy.
More generally, effects beyond those considered in our approximation carry a significant part of
the energy, and call for an extension of our analysis, which is beyond the scope of the present
article.

We observe that the presence of competing ferrimagnetic and antiferromagnetic states with close
energies has been discussed in earlier works~\cite{burkov_prb_2005}, based on numerical simulations
via Quantum Monte Carlo techniques.
Thus, it is possible that the closeness of the energies of ferri- and antiferromagnetic states is a
robust property of the model.
We can thus hypothesize that the near degeneracy found in our simple family of variational states is
directly connected to a close energy spacing in the exact solution to the model.

We note that ferrimagnetic and antiferromagnetic solutions with close energies follow also from
analyses of analytically-tractable variational wavefunctions for the honeycomb quantum dimer
model~\cite{sen_prl_2008, wang_prl_2009}.
Our analysis, besides showing the presence of competing solutions, reveals a low-energy line
connecting them, which we expect to be smooth.

To conclude the discussion, let us compare the results found in this section to those obtained by
different approaches.
As noted above, the classical approximation predicts an energy which is only of order
$J_{xy}^{2}/J_{zz}$ for $J_{xy}\to 0$~\cite{fazekas_pm_1974}, while the projected approximation
gives an energy of the correct order $O(J_{xy})$.
For the finite ratio $J_{xy}/J_{zz} = 0.07$ realized in the KCSO compound, the energy which we find
$e_{xy} = -0.1397 J_{xy}$ is much lower than the classical energy $e_{xy, {\rm cl}} =
-J_{xy}^{2}/[4(J_{xy} + J_{zz})] \simeq -0.016 J_{xy}$.
As shown in Fig.~\ref{landscape2}, the energy which we find is lower than the semiclassical one,
even after spin-wave corrections are included.

The variational energy of the projected spin-product states, instead, is close to the energy
obtained earlier by means of analytically tractable variational wavefunctions for the quantum
dimer model.
In particular, the energy found here is approximately $2 \%$ lower than that of a superfluid
wavefunction, constructed as an equal-amplitude superposition of all Wannier
states~\cite{sen_prl_2008, wang_prl_2009}.
The latter wavefunction, in fact, belongs to the manifold which we considered in this work, as it
can be generated by setting the angles to $\theta_{\rm A}= \theta_{\rm B}= \theta_{\rm C}=\pi/2$.
The energy which we find is lower by approximately $1.4$\% than the variational minimum $e_{xy}
\simeq - 0.13774 J_{xy} $ found in Refs.~\cite{sen_prl_2008, wang_prl_2009} for supersolid
wavefunctions of the quantum-dimer model.

On the other hand, the approximation is still missing a significant fraction (15\%) of the energy
$e_{xy} = -0.160(5) J_{xy}$ reported in Ref.~\cite{kleine_zpb_1992} based on extrapolations of
exact-diagonalization results, showing that effects beyond those considered here are important in
in the energetics of the system.

Finally, let us comment on the in-plane ordered moment.
Our approximation predicts for the optimal canting angle of the ``Y'' structure
$\tilde{\theta}_{\rm
A} \simeq 0.23 \pi$.
A recent preprint~\cite{zhu_arxiv_2026}, reported on experimental measurements of the in-plane
moment, which in our notation imply an angle $\tilde{\theta}_{\rm A}
\simeq 0.18 \pi$.
The values are not too far, supporting the validity of our approach in a first approximation.

\subsection{Magnetization curve}
\label{sec_magnetization_curve}

\begin{figure}[t]
\centering
\includegraphics[scale=1]{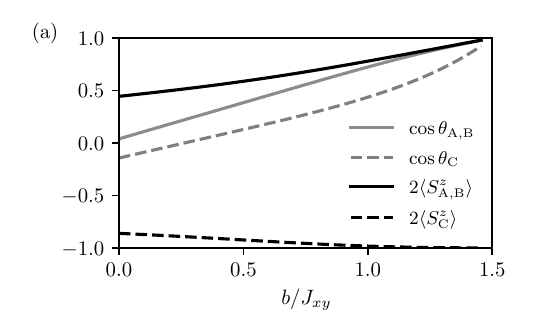}\\
\includegraphics[scale=1]{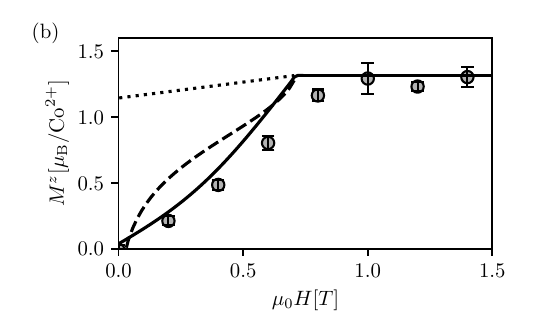}
\caption{\label{magnetization_curve} (a) Variational parameters and longitudinal sublattice
magnetizations as a  function of the magnetic field $b$, computed on an $L= 15$ cylinder. (b)
Theoretical magnetization curve obtained within the projected product-state approximation (solid
line), the classical XXZ model (dotted line), and the classical+ LSWT approximation (dashed line).
Grey circles show experimental data points from Ref.~\cite{zhu_npj_2025}.
The theoretical curves were calculated assuming a gyromagnetic factor $g_{zz} = 7.9$, $J_{xy} =
0.217$ meV, and $J_{zz} = 3.1$meV.
}
\end{figure}

We now turn to an analysis of the magnetization under a longitudinal field.
By analyzing the variational minimum on cylinders of width $L = 12$ we verified that the
lowest-energy state at finite $b$ has the same structure as the ferrimagnetic ground state
($\theta_{\rm A} = \theta_{\rm B}$).
To derive the magnetization curve, we thus analyzed the energy minimum assuming $\theta_{\rm
A}= \theta_{\rm B}$ on cylinders of width $L = 15$.
The results which we obtained are shown in Fig.~\ref{magnetization_curve}.
As the field grows, the average spin directions become more and more aligned to the $z$ axis, until
eventually at the critical field $b_{c} = 3 J_{xy}/2$ the system enters the collinear ``up-up-down''
phase.

As shown in Fig.~\ref{magnetization_curve}a, the bare angles $\theta_{\rm A}$, $\theta_{\rm B}$,
$\theta_{\rm C}$ of the classical state before projection turn out to be substantially
different in comparison with the ``physical'' angles of the projected state, even at large fields.
In particular, $\cos \theta_{\rm C}$ has opposite sign to $\langle S^{z}_{C}\rangle$ at
intermediate and large fields.

In Fig.~\ref{magnetization_curve}b, we present a comparison between the total magnetization 
calculated within the projected coherent state approximation, the semiclassical approximation, and
experimental data for KCSO.
To convert our results to standard units we used $b = g_{zz} \mu_{\rm B} \mu_{0} H$, and assumed 
the parameters $J_{xy} = 0.217$ meV, $g_{zz} = 7.9$ reported in Ref.~\cite{zhu_npj_2025}.

The magnetization curve which we find is close to experimental data, and dramatically improves over
the semiclassical theory.
In particular, the classical theory predicts a magnetization close to $1/3$ for all field
values~\cite{zhu_npj_2025}.
Calculating the magnetization in LSWT via the derivative $-\pa E/\pa b$ of the ground-state energy
we find, consistently with Ref.~\cite{ulaga_prb_2025b},  that the magnetization becomes negative at
low fields, which signals an instability of the spin-wave approximation.

By contrast, the magnetization curve which we find in the projected spin-product
approximation follows satisfactorily the experimental results in a first approximation.

At zero field, we note however that experimental results are consistent with a vanishing
magnetization  $M^{z} = 0$~\cite{chen_nc_2026, zhu_npj_2025}.
The ferrimagnetic state predicted in the projected product-state approximation has a residual
moment, but only a very small one, equal to $\simeq 1$\% of the saturation magnetization.

\subsection{Excitation spectrum at arbitrary momentum}
\label{spectrum_finite_momentum}

\begin{figure*}[t]
\centering
\includegraphics[scale=1]{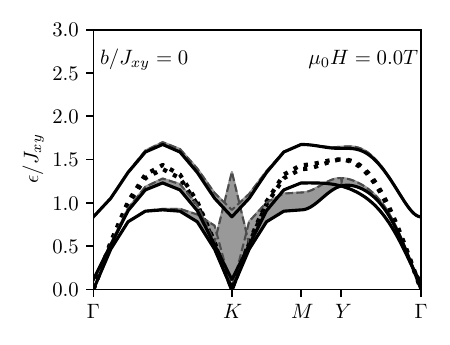}
\includegraphics[scale=1]{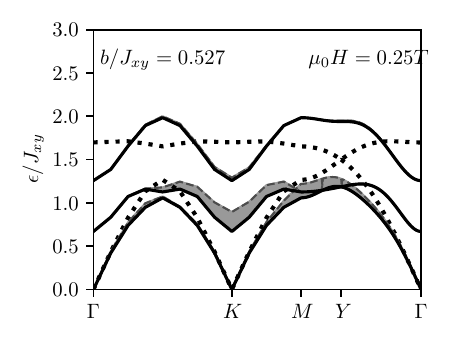}
\includegraphics[scale=1]{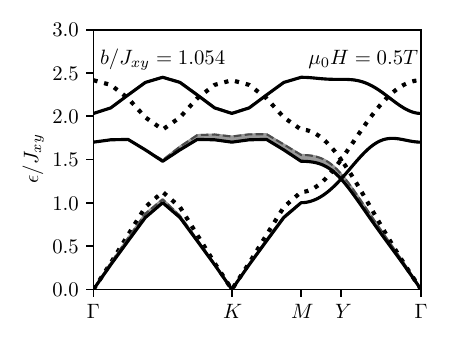}
\caption{\label{e_k_vs_LSWT} Dispersion relations $\epsilon_{n}(\bk)$ and matrix elements $|\langle
n , \bk| S^{z}(\bk)|0\rangle|^{2}$ within the linearized time-dependent variational principle,
calculated on an $L = 12$ cylinder.
The dispersion is represented by solid lines.
The matrix elements of the branches are the dashed lines measured from the corresponding solid
lines.
The dotted curves are dispersions within linear spin-wave theory for $b = g_{zz} \mu_{\rm B} \mu_{0}
H$, $g_{zz} = 7.9$, $J_{xy} = 0.217$ meV, and $J_{zz} = 3.1$ meV.
In the top-left panel, the mode with strong intensity at the $K$ point is the low-energy gapped
pseudo-Goldstone mode.}
\end{figure*}

Finally, we analyze excitations, by using the linearized time-dependent variational principle
approximation.
To calculate the spectrum, we computed numerically the real-space correlation functions $C_{ij}$ and
the stiffness elements $L_{ij}$ on cylinders of width $L=12$ for different values of $b$,
corresponding to the neutron data reported in Ref.~\cite{zhu_npj_2025}.
The correlations $C_{ij}$ and $L_{ij}$ were calculated for all pairs of sites in which $i$ ranges
over $i = (0,0), (1, 0), (2, 0)$, and $j$ ranges over sites with distance $\Delta y = y_{j} - y_{i}
\leq 30$ in the direction parallel to the cylinder axis.
The calculation of $L_{ij}$ requires summations of three-point correlation functions of the form
$\sum_{\langle k, l \rangle }\langle \{ \{(S^{z}_{i} - \langle S^{z}_{i}\rangle ),  (h_{kl} -
\langle h_{kl} \rangle)\},  (S^{z}_{j} - \langle S^{z}_{j}\rangle )\}\rangle$ and $ \langle \{
\{(S^{z}_{i} - \langle S^{z}_{i}\rangle ),  (S^{z}_{k}  - \langle S^{z}_{k}\rangle)\},  (S^{z}_{j} -
\langle S^{z}_{j}\rangle )\}\rangle$, where $h_{kl}$ is the transverse exchange operator acting on
the bond $\langle k, l \rangle$.
These summations were cutoff at a distance $|y_{i} - y_{k}| =  45$ along the cylinder axis.

\begin{figure*}[t]
\centering
\includegraphics[scale=1]{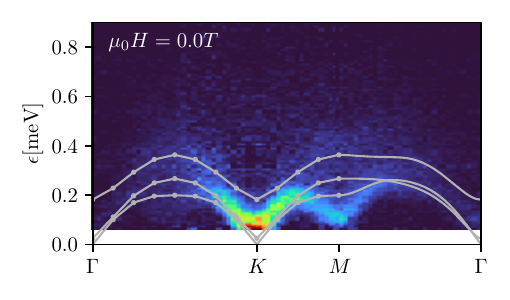}
\includegraphics[scale=1]{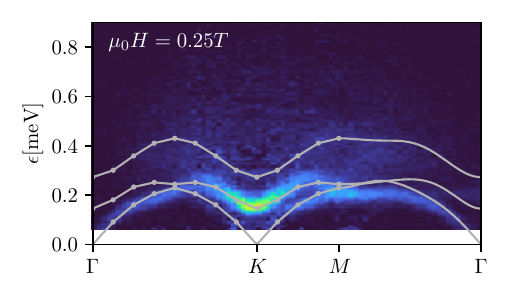} \\
\includegraphics[scale=1]{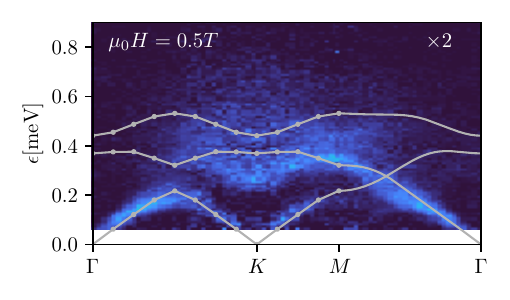}
\includegraphics[scale=1]{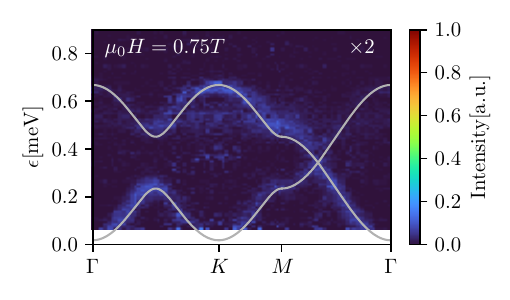}
\caption{\label{INS} Dispersion of the excitation modes within the projected spin-wave approximation
(white lines) for $b = g_{zz} \mu_{\rm B} \mu_{0} H$, $g_{zz} = 7.9$ and $J_{xy} = 0.217$ meV, in
comparison to inelastic neutron scattering data on KCSO from Ref.~\cite{zhu_npj_2025}.
For the spectrum at $H= 0.75$T, we used the analytical expressions $\epsilon_{\bk} = b
\pm \tfrac{1}{2} J_{xy} |f_{\bk}|$, $f_{\bk} = {\rm e}^{i k_{x}} + 2 \cos (\sqrt{3}k_{y}/2) {\rm
e}^{-i k_{x}/2}$ of the dispersion relations in the $1/3$-plateau phase~\cite{zhang_prb_2011,
chen_nc_2026}.
}
\end{figure*}

After determining $C_{ij}$, $L_{ij}$, and the average bond energies the spectrum and the mode
intensities are calculated by Fourier transformation and diagonalization of the dynamical matrix
$D$.
The results which we obtain are shown in Figs.~\ref{e_k_vs_LSWT} and~\ref{INS}, in comparison
with linear-spin-wave theory and neutron-scattering data from Ref.~\cite{zhu_npj_2025}.

The spectrum which we obtain in the linear approximation consists by construction of 3 excitation
branches, since the theory involves linear equations of motion on a background with 3-sublattice
ordering.
Of the three branches which emerge from the diagonalization of $D$, two form a doublet, with a
Dirac point at $Y = 2\pi/3 (1, \sqrt{3}/3)$, and a third branch forms an isolated mode at higher
energy.
The low-energy doublet comprises the gapless Goldstone mode, related to the broken $U(1)$ symmetry,
and a ``pseudo-Goldstone mode'' which has very low energy at small magnetic field.
The third branch, instead, is at higher energy for all field values which we considered.

Because the theory is defined within the subspace of Wannier states, all three branches describe
excitations in the energy range $\epsilon \approx J_{xy}$.
This constrasts with the linear spin wave modes of the non-projected XXZ
model~\cite{kleine_zpb_1992}, which consist of a doublet of low-energy branches with $\epsilon
\approx J_{xy}$, and a separate high-energy branch with $\epsilon \simeq 3J_{zz}$.
The neutron scattering data presented in Refs.~\cite{zhu_prl_2024, chen_nc_2026, zhu_npj_2025} do
not give evidence of a long-lived gapped mode above the lowest-energy states, at energies $\approx
J_{xy}$.
It is likely that the third mode which we find appears as a coherent excitation in the linearized
approximation used here, but is in reality an incoherent superposition of an excitation continuum.
Although this calls for a more detailed analysis, we find that this third branch has a small matrix
element $|\langle n, \bk|S^{z}_{-\bk}| 0\rangle|^{2}$ (see Fig.~\ref{e_k_vs_LSWT}), and contributes
very little to the intensity ${\cal S}^{zz}(\bk, \omega)$.

We thus focus the discussion on the lowest branches.
From Fig.~\ref{e_k_vs_LSWT} we observe that the projected spin-product approximation predicts a
strong reduction of the pseudo-Goldstone gap at the $K$ point in comparison with LSWT, for $H =
0.25$T and $H = 0.5$T.
This leads to a dramatic improvement in the agreement with experimental data, although the gap at
$H=0.5$T is still overestimated by our approximation.

The distribution of the intensities among different momenta and modes also shows a good first
agreement with experimental data.
In particular, we find consistently with experiments that the strongest intensity at low fields is
carried by the pseudo-Goldstone excitations in proximity of the $K$ point.

Despite these improvements there remain discrepancies, revealing the importance of correlations
beyond those considered in this work.
Most importantly, the approximation used here does not explain the appearance of roton-like minima
at the $M=\pi(1, \sqrt{3}/3)$ points and at the points $K/2=(2\pi/3, 0)$~\cite{chen_nc_2026,
zhu_npj_2025}.
We observe that $M$ and $K/2$ are related by a reciprocal vector, after folding the Brillouin zone
on account of the three-sublattice ordering.
Thus, the minima at these points can be traced to common origins.
The linearized TDVP theory does not generate minima but rather saddle points, with a local minimum
in the $\Gamma-M$ directions and a local maximum in the $\Gamma-K$ directions.
The description of the mimima thus remain beyond the approximation considered here.

\subsection{Pseudo-Goldstone gap at the $K$ point}
\label{excitation_spectrum_pG}

We now focus more closely on the spectrum at the $K$ point.
At this point of the Brillouin zone, neutron scattering measurements on KCSO revealed excitations
with a strong intensity and a small gap $\epsilon_{\rm g} \simeq 60\mu$eV$\simeq 0.3 J_{xy}$ at zero
field~\cite{chen_nc_2026, zhu_npj_2025, zhu_prl_2024}, evolving into a larger-gap excitation at
higher fields.

This mode was interpreted as a ``pseudo-Goldstone'' mode~\cite{miyashita_jpsj_1985, murthy_prb_1997,
chen_nc_2026, lin_arxiv_2025, mauri_prb_2026}.
It has been suggested~\cite{zhu_npj_2025} that its low energy can be traced to the degeneracy of
the classical XXZ model at zero magnetic field, which leads to a ``soft'' deformation of the ground
state.

The analysis shown in Sec.~\ref{zero_field_ground_state} showed that the explicit consideration of
a projected model is essential in the limit $J_{xy} \to 0$ and that the projection into the
subspace of Wannier states dramatically changes the wavefunctions relative to their classical
counterparts.
Nevertheless, the qualitative prediction of a soft mode remains valid in the projected spin-product
approximation: deforming the ground state along the flat direction of the energy landscape has very
low energy cost.

To connect the gap of the $K$ point mode to the curvature of the energy, we can take advantage of
the fact that the spectrum at the $K$ point can be derived by studying perturbations which preserve
the three-sublattice periodicity.
Since the pseudo-Goldstone fluctuation, experimentally, has intensity only at $K$ point and not at
the $\Gamma$ point, it can be identified with an antisymmetric mode in which $\delta \theta_{\rm C}
= \delta \varphi_{\rm C} = 0$, $\delta \theta_{\rm A} = - \delta \theta_{\rm B}$,
$\delta \varphi_{\rm A} = -\delta \varphi_{\rm B}$.
Introducing $\varphi_{\rm antisym}= \sqrt{N}(\varphi_{\rm A} - \varphi_{\rm B})/\sqrt{6}$,
$m^{z}_{\rm antisym} = \sqrt{N} (\langle S^{z}_{\rm A}\rangle - \langle S^{z}_{\rm
B}\rangle)/\sqrt{6}$ and analyzing the dynamics of this normal mode via Eqs.~\eqref{eom1} we find
that the frequency of the antisymmetric fluctuation can be expressed in the form:
\begin{equation} \label{pseudo_Goldstone_gap}
\begin{split}
\epsilon_{\rm g} & = \sqrt{\frac{\pa^{2}E}{\pa^{2} m_{\rm antisym}^{z2}}
\frac{\pa^{2}E}{\pa \varphi_{\rm antisym}^{2}}} = \sqrt{{\cal C}_{m} {\cal C}_{\varphi}}~.
\end{split}
\end{equation}

We thus have, in the linear approximation, a direct relation between the curvature of the energy
function and the gap (which is expected because of the harmonic nature of the approximation
considered).
The curvature ${\cal C}_{\varphi}$ can be directly related to the average energies $e_{\rm AB}$ and
$e_{\rm BC}$ of A-B and B-C bonds and can be shown to equal ${\cal C}_{\varphi} = - 6 e_{\rm AB} -
3 e_{\rm BC}$.
${\cal C}_{m}$ instead measures the curvature of the energy in the direction tangent to the
quasi-degenerate ground state manifold, locally, in an infinitesimal neighbourhood of the ``Y''
ground state.

To determine quantitatively ${\cal C}_{m}$ we calculated the second-derivative of the energy in a
finite-difference approximation, by applying a small antisymmetric perturbation to the ground-state
solution.
In this way we find for zero field ${\cal C}_{\varphi} \simeq 0.6 J_{xy}$,  ${\cal C}_{m} \simeq
0.024 J_{xy}$, and $\epsilon_{\rm g} \simeq 0.12 J_{xy}$.

Instead of the gapless pseudo-Goldstone mode predicted by LSWT we thus have a small, but finite
gap, in the projected approximation.

In comparison with experimental results on KCSO $\epsilon_{\rm g} \simeq 0.3 J_{xy}$, the value of
the gap is too small, indicating that a consideration of further correlations beyond the mean-field
one is essential.
An extension of our analysis is a challenging problem beyond our present scope.

We remark however that the explicit consideration of a projected theory is essential in this
direction.
In the non-projected $1/S$ expansion, indeed, the limit $J_{xy} \to 0$ produces an anomalously
large pseudo-Goldstone gap, which overestimates the experimental value by more than an order of
magnitude~\cite{mauri_prb_2026}.
The large value of $\epsilon_{\rm g}$ follows from large corrections occurring in the $1/S$
expansion near the Ising limit.
The projected theory considered here automatically avoids these difficulties, and provides a
starting point towards a further analysis which does not suffer from corrections which are
parametrically large in $J_{zz}/J_{xy}$.

\begin{figure}
\includegraphics[scale=1]{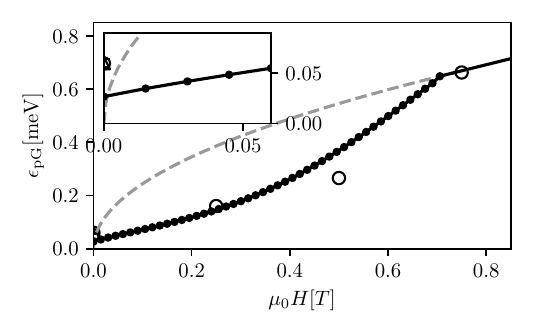}
\caption{\label{pG_gap} Pseudo-Goldstone gap as a function of the longitudinal field $b$ within the
projected linear-spin-wave approximation on cylinders of width $L = 12$ (black solid line), and
within LSWT (gray dashed line).
The energy and magnetic field scales were set by assuming $J_{xy} =0.217$meV, $g_{zz} = 7.9$.
The inset shows the same data, enlarging the region near $b=0$.
The experimental data points marked with a circle were extracted from Ref.~\cite{zhu_npj_2025}.
At $b= 0$, Refs.~\cite{zhu_npj_2025} and~\cite{chen_nc_2026} reported consistent values of the gap,
$\simeq 60 \mu$eV.
For $b > b_{c}$ the gap is $\epsilon_{\rm g} = (b - b_{c}) + \epsilon_{\rm g}(b_{c})$.}
\end{figure}

To further analyze the pseudo-Goldstone mode at the $K$ point, we calculated its evolution with $b$
by evaluating numerically Eq.~\eqref{pseudo_Goldstone_gap}.
The results are presented in Fig.~\ref{pG_gap} in comparison with with neutron scattering data,
which we extracted from Ref.~\cite{zhu_npj_2025}, and with the LSWT prediction for $J_{xy}/J_{zz}
\to 0$, $\epsilon_{\rm g} = \sqrt{6 b J_{xy}}$~\cite{kleine_zpb_1992}.
As it can be seen, the projected approach predicts a value of the gap $\epsilon_{\rm g}$ which is
significantly reduced at intermediate fields in comparison with LSWT, and is closer to experimental
data.

\section{Conclusions}
\label{conclusions}

In conclusion, we have introduced and analyzed an approximate theory for the strongly anisotropic
quantum XXZ model, based on projected spin-product states, which are analogue to classical
spin-states, but projected within the low-energy subspace of Wannier configurations.
Our analysis showed that the model presents at zero field nearly degenerate ground states, and a
flat a one-dimensional line of quasi-degenerate minima connecting competing ferrimagnetic and
antiferromagnetic ground states.
Via a time-dependent variational principle, we connected this finding to the presence of a
low-energy pseudo-Goldstone mode in the spectrum.

The magnetization curve predicted by our approach presents a good first agreement with experimental
data reported for K$_{2}$Co(SeO$_{3}$)$_{2}$.
The excitation spectrum computed in our approximation shows a significantly improved agreement in
comparison with the linear-spin-wave approximation.
Besides, the explicit consideration of a projected model avoids large corrections encountered in
the $1/S$ expansion in the case in which $J_{xy}$ is much smaller than $J_{zz}$.

There remain aspects of the experimental data of KCSO which require a more detailed theory.
In particular, the pseudo-Goldstone gap predicted by our approximation at zero field is too small
in comparison with the experimental value.
Secondly, the linearized equations of motion do not capture the presence of roton-like minima at
the $M$ and the $K/2$ points.

The theory pressented here provides a first step towards an analysis of the complex spectrum of the
triangular quantum Ising model within a ``spin-wave'' picture, while maintaining an exact
enforcement of the Ising- or, equivalently, quantum-dimer constraints.

\vspace{\baselineskip}
This work has been supported by the Swiss National Science Foundation Grant No. 212082.
We are grateful to Andrey Zheludev and Collin Broholm for useful discussions and to Andrey Zheludev
and Mengze Zhu for providing experimental data for K$_{2}$Co(SeO$_{3}$)$_{2}$.

\appendix

\section{Ground-state averages in projected coherent states}
\label{averages}

The probability distribution of spin configurations $|\psi(\sigma)|^{2}$ within a projected
coherent state is equivalent to the distribution of a classical Ising model with $\beta H =
\sum_{\langle i, j \rangle} \lt(J_{zz} S^{z}_{i} S^{z}_{j} - h_{i} S^{z}_{i}\rt)$, in the
limit of infinite $J_{zz}$, under the action of a space-dependent magnetic field field $h_{i} = \ln
\lt((1 + \cos \theta_{i})/(1 - \cos \theta_{i})\rt)$.
All averages and correlation functions can thus be expressed in terms of averages within the
corresponding Ising model.

Averages involving only products of $S^{z}_{i}$ operators can be calculated directly from the
distribution $|\psi(\sigma)|^{2}$.
Averages involving transverse operators $S^{x}_{i}$, $S^{y}_{i}$, instead, require additional phase
factors, which can be simply computed.

\begin{figure}[h]
\centering
\includegraphics[scale=1]{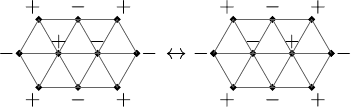}
\caption{\label{interchangeable_pair} Due to the projection onto the subspace of Ising ground
states, a spin pair can be flipped by the transverse operator $P(S^{x}_{i} S^{x}_{j} + S^{y}_{i}
S^{y}_{j})P$ only if it is surrounded by 8 spins with alternating signs.
Flipping a spin pair which is not surrounded by alternating spins would lead to a violation of the
Ising ground-state rule and the corresponding processes are thus annihilated by the projector $P$.}
\end{figure}

The projected transverse-exchange interaction $P (S^{x}_{i} S^{x}_{j} + S^{y}_{i} S^{y}_{j})P$ has
the effect of flipping a pair of nearest-neighbour spins $i$ and $j$ if the two spins are
surrounded by a shell of 8 alternating ``up'' and ``down'' spins (see
Fig.~\ref{interchangeable_pair})~\cite{fazekas_pm_1974, moessner_prb_2001, sen_prl_2008,
wang_prl_2009}.
Spin-pairs which are not surrounded by a shell of alternating spins, instead, cannot be flipped
without leaving the subspace of Ising states, and do not contribute to the average $\langle
P(S^{x}_{i} S^{x}_{j} + S^{y}_{i} S^{y}_{j})P\rangle$.
The average transverse-exchange energy thus involves 8-spin correlations.

By a direct calculation we find that the average energy of a single nearest-neighbour bond $e_{ij}$
within a projected spin-product state, can be expressed as:
\begin{equation}
\begin{split}
e_{ij} & = -J_{xy} \sin \theta_{i} \sin \theta_{j} \cos (\varphi_{i} -
\varphi_{j}) \\
& \times \Bigg(\frac{P_{+}}{3 + \cos \theta_{i}
+ \cos \theta_{j} - \cos \theta_{i} \cos \theta_{j}} \\
& + \frac{P_{-}}{3 - \cos \theta_{i} - \cos \theta_{j} - \cos \theta_{i} \cos \theta_{j}} \Bigg)
~.
\end{split}
\end{equation}

\begin{figure}[h]
\centering
\includegraphics[scale=1]{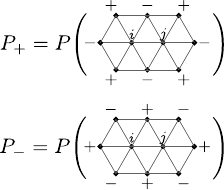}
\caption{\label{probabilities} Probabilities $P_{+}$, $P_{-}$ of the two alternating configurations
for which the internal pair is interchangeable.
The configuration '$+$' is such that $\sigma^{z}_{i}$ and $\sigma^{z}_{j}$ can be both $+$, but
cannot be both $-$.
The configuration '$-$' is such that $\sigma^{z}_{i}$ and $\sigma^{z}_{j}$ can be both $-$, but
cannot be both $+$.
$P_{+}$ and $P_{-}$ are the total probabilities of the 8-spin configurations '$+$' and '$-$',
summed over the possible configurations of the internal spin pair.}
\end{figure}

Here, $P_{+}$ and $P_{-}$ are the probabilities to find the two 8-spin configurations shown in
Fig.~\ref{probabilities}, calculated with the distribution $|\psi(\sigma)|^{2}$.

\begin{figure}[h]
\centering
\includegraphics[scale=1]{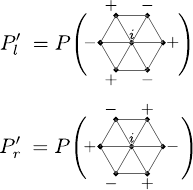}
\caption{\label{probabilities_Sx} Probabilities $P'_{l}$, $P'_{r}$ of the two alternating
configurations for which the spin is flippable.
The configuration '$l$' is defined as the configuration having a negative spin to the left of site
$i$. The configuration '$r$' has a negative spin to the right of site $i$.
$P'_{+}$ and $P'_{-}$ are defined as the total probabilities of the six-spin configurations '$+$'
and '$-$', summed over the two possible orientation of the internal spin.}
\end{figure}

The operators $P S^{x}_{i} P$, $PS^{y}_{i}P$ have the effect of flipping a single spin at the site
$i$, if the six spins surrounding it have alternating signs.
The average of the transverse moments in a projected coherent state can be expressed as
\begin{equation}
\begin{split}
&\langle P S^{x}_{i} P \rangle = \frac{1}{2} \sin \theta_{i} \cos \varphi_{i} \lt(P'_{r} +
P'_{l}\rt)~, \\
& \langle P S^{y}_{i} P \rangle = \frac{1}{2} \sin \theta_{i} \sin \varphi_{i} \lt(P'_{r} +
P'_{-}\rt)~,
\end{split}
\end{equation}
where $P'_{l}$ and $P'_{r}$ are the probabilities to find the two configurations in
Fig.~\ref{probabilities_Sx}, computed with the distribution $|\psi(\sigma)|^{2}$.

To calculate averages in the AFM model with $J_{xy} > 0$, as discussed in the main text, we
apply the unitary transformation $U$ introduced in Ref.~\cite{wang_prl_2009}, to the variational
state which optimizes the energy of the $J_{xy} < 0$ problem.
The action of $U$ on $PS^{x}_{i}P$ and $P S^{y}_{i} P$ can be expressed as:
\begin{equation}
\begin{split}
U^{+} PS^{x}_{i} P U = 2 S^{z}_{i - \hat{x}} S^{y}_{i}~, \\
U^{+} PS^{y}_{i} P U = - 2 S^{z}_{i - \hat{x}} S^{x}_{i}~, \\
\end{split}
\end{equation}
where $i-\hat{x}$ is the site immediately to the left of site $i$.
Applying for convenience also a global $\pi/2$ rotation we obtain the expressions for the in-plane
moments in the AFM state:
\begin{equation}
\begin{split}
&\langle P S^{x}_{i} P \rangle_{\rm AFM} = \frac{1}{2} \sin \theta_{i} \cos \varphi_{i}
\lt(P'_{r} - P'_{l}\rt)~,\\
&\langle P S^{y}_{i} P \rangle_{\rm AFM} = \frac{1}{2} \sin \theta_{i} \sin \varphi_{i}
\lt(P'_{r} - P'_{l}\rt)~.\\
\end{split}
\end{equation}

Correlations involving both bond-energy and spin operators were determined by analogue calculations.

\section{Details of the numerical transfer-matrix calculations}

\begin{figure}[h]
\centering
\includegraphics[scale=1]{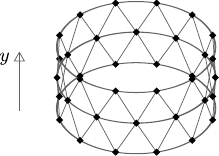}
\caption{\label{transfer_matrix} Geometry of the cylinders used in numerical calculations.}
\end{figure}

To compute numerically averages with the distribution $|\psi(\sigma)|^{2}$, we used a transfer
matrix method, on cylinders of infinite length and finite width $L  = 3n$.
We used cylinders in which the $y$ axis (infinite direction) is orthogonal to the bonds, and the
$x$ axis (the compact dimension) is parallel to the bonds, as shown in Fig.~\ref{transfer_matrix}.
In the calculation of the spectrum, this gives a momentum resolution $\Delta k_{x} = 2\pi/L$
along the $\Gamma-K$ axis.
The momentum resolution along the $y$ axis, instead, is virtually infinite, because of the infinite
extent of the cylinder in the $y$ direction.

In order to calculate expectation values on infinite-length cylinders, we calculated via exact
diagonalization the maximum eigenvalue $\lambda_{\rm max}$, and the corresponding eigenvector
$v_{\rm
max}$ of the even-row to even-row transfer matrix, \emph{i.e.} the transfer matrix connecting the
row at $y_{n} = \tfrac{\sqrt{3}}{2} n$ to the row at $y_{n} = \tfrac{\sqrt{3}}{2} (n+2)$, with $n$
even.
The probability to find a specified set of spin values $\sigma^{z}_{1}$, ...., $\sigma^{z}_{M}$
on a set of sites $i_{1}$, ..., $i_{M} = (x_{1}, y_{1}), ..., (x_{M}, y_{M})$ was then calculated
by contracting matrix products of the form:
\begin{equation}
\begin{split}
& P_{i_{1}, ... i_{M}}(\sigma^{z}_{1} , ..., \sigma^{z}_{M}) = \lambda_{\rm max}^{-(h_{1} - h_{0})}
v^{T}_{\rm max} (\\
& \qquad t_{\rm odd} \Lambda_{h_{1} - 1} t_{\rm even}\Lambda_{h_{2}-1} ... \\
& \qquad \Lambda_{h_{0}+2} t_{\rm odd}\Lambda_{h_{0}+1} t_{\rm even}\Lambda_{h_{0}}) v_{\rm max}~.
\end{split}
\end{equation}

In this expression, $h_{0}$ and $h_{1}$ are even integers respectively smaller and larger than all
the coordinates $2 \sqrt{3}y_{i}/3$, involved in the correlation function.
$t_{\rm even}$ and $t_{\rm odd}$ are transfer matrices connecting even to odd and odd to even rows.
$\Lambda_{i}$ are diagonal transfer matrices equal to $1$ for configurations consistent with the
spin-configuration $\sigma^{z}_{1}$, ...., $\sigma^{z}_{M}$ and $0$ otherwise.

\begin{figure}[h]
 \centering
\includegraphics[scale=1]{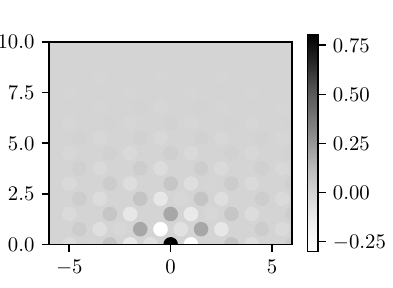}\\
 \includegraphics[scale=1]{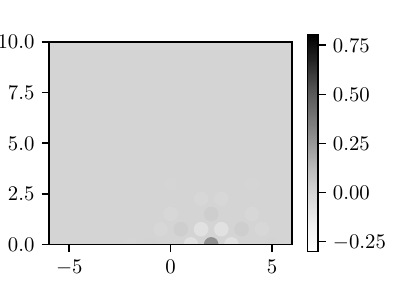}
  \caption{\label{c} Connected correlation $4C_{ij} = 4 \langle (S^{z}_{i} - \langle
S^{z}_{i} \rangle) (S^{z}_{j} - \langle S^{z}_{j} \rangle) \rangle$ as a function of $j$ for a
fixed $i$ in the A sublattice (top panel) and in the C sublattice (bottom panel).}
\end{figure}

\begin{figure}[h]
 \centering
\includegraphics[scale=1]{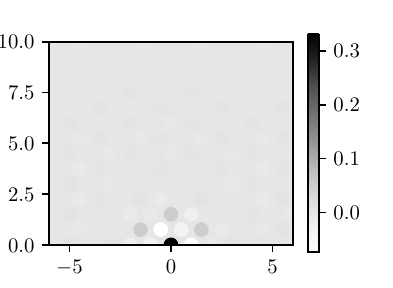} \\
\includegraphics[scale=1]{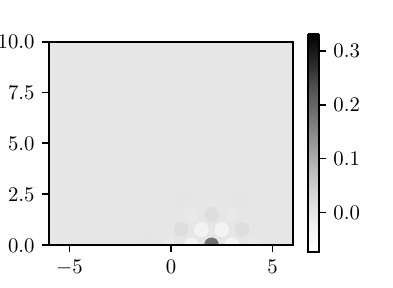}
  \caption{\label{l} Real-space susceptibility matrix $L_{ij}$ as a function of $j$ for a fixed $i$
in the A sublattice (top panel) and in the C sublattice (bottom panel).}
\end{figure}

Figs.~\ref{c} and~\ref{l} show, respectively,  the real space correlation function $C_{ij}$ and the
matrix $L_{ij}$ which we obtained on a cylinder of width $L = 12$ and for $b=0$.

From Fig.~\ref{l} one may expect that $L_{ij}$ decreases quickly with distance, and that its
Fourier transform should only suffer from a small error when summations over $j$ are truncated to a
small distance.
Contrary to this expectation, in our numerical calculation we found that the Fourier transform
converges rather slowly at low field, and it was necessary to consider large cutoff distances
($\Delta y \simeq 30$).

We verified convergence with respect to the cutoff along the vertical direction, by comparing the
zero-momentum Fourier components $L_{\alpha \beta}(\bk = 0)$ with the corresponding expressions
$L_{\alpha \beta} = \tfrac{3}{2 N} \sin \theta_{\alpha} \sin \theta_{\beta} \frac{\pa^{2} E}{\pa
\theta_{\alpha} \pa \theta_{\beta}}$, $\alpha, \beta $= A, B, C.
Calculating the latter derivatives by a finite-difference approximation, we verified agreement of
the zero-momentum transform within $2$ \% for $L_{01}(\bk = 0)$ component at zero field, and with a
higher precision ($\approx 0.1$ \%) for other matrix elements and higher fields.

\section{Matrix elements for the ${\cal S}^{zz}$ structure factor}
\label{dynamical_structure_factor}

To extract the spectral intensity, we approximate the linear-response function by
considering linear perturbations within the TDVP framework, including a source term $-\sum_{i}
J_{i}(t) S^{z}_{i}$.
The source introduces an inhomogeneous term into the equations of motion~\eqref{eom1}, which become:
\begin{equation} \label{eom3}
\begin{split}
& \sum_{j} \lt(   L_{ij} x_{j} + 2 C_{ij} (\dot{\varphi_{j}} +J_{j}(t))\rt) = 0~.\\
&\sum_{j} \lt(2 C_{ij} \dot{x}_{j} - M_{ij} \varphi_{j} \rt) = 0~, \\
\end{split}
\end{equation}

Inverting the kernel, the susceptibility in real space and in the frequency domain can be expressed
as
\begin{equation} \label{chi_zz}
\begin{split}
& \chi^{zz}_{ij}(\omega) = \frac{\pa \delta \langle S^{z}_{i}\rangle(\omega)}{\pa J_{j}(\omega)}
\\
& = 4 \begin{vmatrix}
C^{-1} L C^{-1} & -2 i (\omega+i0^{+}) \\
2 i (\omega+i0^{+}) & M
\end{vmatrix}_{11, ij}^{-1} \\
& = 4 \lt[A^{-1}(\omega) \rt]_{11, ij}~.
\end{split}
\end{equation}

Here $A_{11}$ denotes the top-left $N \times N$ corner of the $(2N)\times (2N)$-dimensional matrix
$A(\omega)$.

$A$ can be inverted in the basis of its eigenvectors.
Due to translational symmetry, the eigenvectors of $A$ are plane waves with momentum $\bk$.
We can write the $2N$-dimensional eigenvectors of $A$ as $v_{n}(\bk) = (v_{n, 1, i}(\bk), v_{n, 2,
i}(\bk))$, where $v_{n, \alpha, i}(\bk, \omega) = \sqrt{3/N} {\rm e}^{i \bk \cdot \bx_{i}}
\tilde{v}_{n, \alpha, i}(\bk)$, $\alpha = 1, 2$, and the amplitude $\tilde{v}_{n, \alpha, i}(\bk)$
depends on $i$ only via its sublattice (A, B, or C).

In the $v_{n}$ basis the susceptibilities in real and momentum space can thus be expressed as:
\begin{equation}
\chi^{zz}_{ij}(\omega) = \frac{12}{N} \sum_{n=1}^{6}\sum_{\bk} {\rm e}^{i \bk \cdot (\bx_{i} -
\bx_{j})}\frac{\tilde{v}_{n, 1, i}(\bk, \omega) \tilde{v}^{*}_{n, 1, j}(\bk, \omega)}{v^{+}_{n}(\bk,
\omega) A(\omega) v_{n}(\bk, \omega)}~,
\end{equation}
\begin{equation}
\chi^{zz}(\bk, \omega) = \frac{4}{3} \sum_{n=1}^{6} \frac{ \lt|  \sum_{j={\rm A},
{\rm B}, {\rm C}}\tilde{v}_{n, 1, j}(\bk, \omega)\rt|^{2}}{v^{+}_{n}(\bk,
\omega) A(\omega) v_{n}(\bk, \omega)}~.
\end{equation}

It can be verified that $\det A(\omega) = 0$ when $\omega$ is equal to one of the spectral
frequencies.
Thus $A(\omega)$ has poles coinciding with the excitation frequencies.
The imaginary part of $\chi$, which determines the dynamical structure factor, arises because of
the poles at $\omega = \epsilon_{n}(\bk)$, and can be
calculated by substituting $\omega \to \omega + i0^{+}$ and by using the Dirac relation ${\rm Im}
1/(x+i0^{+}) = -\pi \delta(x)$.
Using $A(\epsilon_{n}(\bk)) v_{n}(\bk, \epsilon_{n}(\bk)) = 0$ and
\begin{equation}
\frac{\pa A(\omega)}{\pa \omega} = 2
\Sigma_{y} = 2 \begin{vmatrix}
                0_{N\times N} & -i\times I_{N \times N} \\
                i \times I_{N\times N} & 0_{N\times N}
               \end{vmatrix}~,
\end{equation}
the residue can be extracted by replacing $v^{+}_{n}(\bk, \omega) A(\omega)
v_{n}(\bk, \omega) \simeq 2(\omega - \epsilon_{n}(\bk) + i 0^{+}) (v^{+}_{n}(\bk, \epsilon_{n}(\bk))
\Sigma_{y} v_{n}(\bk, \epsilon_{n}(\bk)))$.

From the Dirac relation we then find:
\begin{equation}
\begin{split}
{\cal S}^{zz}(\bk, \omega) & = -\frac{2}{3}  \Theta(\omega) \sum_{n=1}^{6}  \frac{\lt|\sum_{j={\rm
A}, {\rm B}, {\rm C}}\tilde{v}_{n, 1, j}(\bk, \epsilon_{n}(\bk))\rt|^{2}}{v^{+}_{n}(\bk,
\epsilon_{n}(\bk)) \Sigma_{y} v_{n}(\bk, \epsilon_{n}(\bk))}  \\
& \qquad \times \delta (\omega - \epsilon_{n}(\bk)) \\
\end{split}
\end{equation}

The expression can be rewritten conveniently in terms of the right-hand eigenvectors $u_{n}(\bk)$ of
the dynamical matrix $D(\bk)$ in momentum space.
Denoting as $(u_{n, 1, i}(\bk)$, $u_{n, 2, i}(\bk))$, $i = {\rm A}, {\rm B}, {\rm C}$ the
eigenvector with eigenvalue $\epsilon_{n}(\bk)$, it can be checked that: $\tilde{v}_{n, 1, i}(\bk,
\epsilon_{n}(\bk)) = u_{n, 2, i}(\bk)$,  $\tilde{v}_{n, 2, i}(\bk, \epsilon_{n}(\bk)) = u_{n, 1,
i}(\bk)$.
The spectrum of $D(\bk)$ is composed of three pairs of equal and opposite eigenvalues.
For $\omega > 0$, only the positive energy eigenvalues contribute.
Thus the expression can be rewritten as:
\begin{equation}
\begin{split}
{\cal S}^{zz}(\bk, \omega) & = \frac{2}{3} \sum_{n=1}^{3}  \frac{\lt|\sum_{j={\rm A}, {\rm B}, {\rm
C}} u_{n, 2, j}(\bk)\rt|^{2}}{u^{+}_{n}(\bk) \Sigma_{y} u_{n}(\bk)} \\
& \qquad \times \delta (\omega - \epsilon_{n}(\bk))~,
\end{split}
\end{equation}
where the sum runs over positive-energy modes.
Using the eigenvalue equation $D(\bk) u_{n}(\bk) = \epsilon_{n}(\bk) u_{n}(\bk)$ and the explicit
form of the matrix $D$ we find that the expression can be eventually recast as:
\begin{equation}\label{Szz}
\begin{split}
{\cal S}^{zz}(\bk, \omega) & = \frac{2}{3} \sum_{n=1}^{3}  \frac{\epsilon_{n}(\bk) \lt|\sum_{j={\rm
A}, {\rm B}, {\rm C}} u_{n, 2, j}(\bk)\rt|^{2}}{u^{+}_{n, 2}(\bk) \tilde{l}(\bk) u_{n, 2}(\bk)} \\
& \qquad \times \delta (\omega - \epsilon_{n}(\bk))~,
\end{split}
\end{equation}
where $\tilde{l}(\bk)$ is the sublattice-resolved stiffness in momentum space:
\begin{equation}
\tilde{l}_{\alpha \beta}(\bk) = \frac{3}{N} \sum_{i \in \alpha, j \in \beta} {\rm e}^{-i \bk \cdot
(\bx_{i} - \bx_{j})} \lt[C^{-1} L C^{-1} \rt]_{ij}~.
\end{equation}

From Eq.~\eqref{Szz} we identify the matrix elements:
\begin{equation}
\begin{split}
&|\langle n, \bk |S^{z}_{-\bk}| 0 \rangle|^{2} = \frac{1}{N} \lt |\sum_{i}\langle n, \bk |{\rm e}^{i
\bk \cdot x_{i}} S^{z}_{i} | 0\rangle\rt|^{2} \\
& \qquad =  \frac{2\epsilon_{n}(\bk) \lt|\sum_{i={\rm A}, {\rm B}, {\rm C}} u_{n, 2,
i}(\bk)\rt|^{2}}{3 u^{+}_{n, 2}(\bk) \tilde{l}(\bk) u_{n, 2}(\bk)}
\end{split}
\end{equation}

We use this expression to calculate the intensity of the three excitation modes in
Fig.~\ref{e_k_vs_LSWT}.

\end{document}